\documentclass[english,aps,manuscript]{revtex4}
\usepackage[T1]{fontenc}
\usepackage[latin9]{inputenc}
\usepackage[letterpaper]{geometry}
\usepackage{color}
\usepackage{textcomp}
\usepackage{amsthm}
\usepackage{amsmath}
\usepackage{esint}

\makeatletter
\@ifundefined{textcolor}{}
{%
 \definecolor{BLACK}{gray}{0}
 \definecolor{WHITE}{gray}{1}
 \definecolor{RED}{rgb}{1,0,0}
 \definecolor{GREEN}{rgb}{0,1,0}
 \definecolor{BLUE}{rgb}{0,0,1}
 \definecolor{CYAN}{cmyk}{1,0,0,0}
 \definecolor{MAGENTA}{cmyk}{0,1,0,0}
 \definecolor{YELLOW}{cmyk}{0,0,1,0}
 }
\numberwithin{equation}{section}
\numberwithin{figure}{section}

\makeatother

\usepackage{babel}

\begin{document}

\title{{\huge Entropy production in relativistic binary mixtures}}

\author{Valdemar Moratto$^{1}$, A. L. Garcia-Perciante$^{2}$, L. S. Garcia-Colin$^{1\,\text{and}\,3}$}

\address{$^{1}$Depto. de Fisica, Universidad Autonoma Metropolitana-Iztapalapa,
Av. Purisima y Michoacan S/N, Mexico D. F. 09340, Mexico.}

\address{$^{2}$Depto. de Matematicas Aplicadas y Sistemas, Universidad Autonoma
Metropolitana-Cuajimalpa, Artificios 40 Mexico D.F 01120, Mexico.}

\address{$^{3}$El Colegio Nacional, Luis Gonzalez Obregon 23, Centro Historico,
Mexico D. F. 06020, Mexico.\\
 }
\begin{abstract}
In this paper we calculate the entropy production of a relativistic
binary mixture of inert dilute gases using kinetic theory. For this
purpose we use the covariant form of Boltzmann's equation which, when
suitably transformed, yields a formal expression for such quantity.
Its physical meaning is extracted when the distribution function is
expanded in the gradients using the well-known Chapman-Enskog method.
Retaining the terms to first order, consistently with Linear Irreversible
Thermodynamics we show that indeed, the entropy production can be
expressed as a bilinear form of products between the fluxes and their
corresponding forces. The implications of this result are thoroughly
discussed. 
\end{abstract}
\maketitle

\section{Introduction}

In a recent paper \cite{Val-3} we have studied two important thermodynamic
aspects of a relativistic binary mixture of inert dilute gases using
the principles of kinetic theory. The first one concerns the so called
cross effects, in this case when local thermal equilibrium is assumed,
they are the well known Dufour and Soret effects \cite{Groot Mazur}.
The second and most relevant one, concerns with the validity of the
Onsager Reciprocity Relations (ORR). As we showed in that paper they
hold true in two representations, or choices, of fluxes and forces.
In the first representation, which is referred to it in the literature
\cite{Groot-Leewen-Weert,CERCIGNANI} the heat flux is coupled to
a modified Fourier-like force involving both, temperature and pressure
gradients. In such representation however, the Dufour and Soret effects
do not appear in their canonical form. The second representation is
rather singular. Introducing the concept of a {}``volumetric flow''
which arises from the relativistic non-invariance of the volume elements
in the fluid and whose force turns out to be the pressure gradient,
the ORR are shown to hold true and the canonical form of such effects
is recovered. Further, this representation is strictly valid only
in the relativistic case.

As it is well-known in Linear Irreversible Thermodynamics (LIT) \cite{Groot Mazur}
the appropriate choice of forces and fluxes is strongly suggested
by the entropy balance equation which in kinetic theory arises by
simply multiplying Boltzmann's equation by the logarithm of the single
particle distribution function and averaging over all the velocities
of the particles. This procedure allows one to identify the entropy
production, usually denoted by $\sigma$ as a bilinear form of products
between forces and fluxes. Symbolically, if $X_{i}$ is the force
associated with a flux $J_{i}$,\begin{equation}
\sigma=\sum_{i}X_{i}\odot J_{i}\label{eq:0}\end{equation}
where, for an isotropic system, $\odot$ denotes full contraction
of the two tensors, necessarily of the same rank.

The full derivation of Eq. (\ref{eq:0}) for a relativistic binary
mixture of inert, diluted gases is the main objective of this paper.
The outcome of this derivation should provide full support for the
force-flux representations used in \cite{Val-3}. Since the whole
scheme of both papers is restricted to the tenets of LIT we will only
need to carry out the calculation using the well-known Chapman-Enskog
expansion to first order in the gradients. As shown in Ref. \cite{Ana 2008}
this procedure is perfectly valid in a relativistic framework characterized
by Minkowski's metric namely, in special relativity. 

To facilitate the reading of this paper we use the following conventions.
Tensors with space and time components are labeled with greek subscripts:
$\alpha,\beta=1,2,3,4$, where the fourth component refers to time
whereas spatial components are labeled with latin subscripts $i,j=1,2,3$.
Einstein's sum convention is adopted for both types of subscripts
throughout, but not for sums characterizing the species of the mixture.
The Minkowski metric has signature $\left\{ +++-\right\} $, colons
and semicolons denote ordinary and covariant derivatives respectively.

The structure of this paper is as follows: In section 2 the basic
ideas of relativistic kinetic theory are explained up to a linearized
version of the Boltzmann equation consistent with the Chapman-Enskog
expansion. The solution to such equation is proposed in section 3
using the representation introduced in Ref. \cite{Val-3}. The general
structure of the entropy balance equation and thus the expression
for the entropy production $\sigma$ are also shown in section 3.
Section 4 is devoted to establishing the required form of $\sigma$
within this representation (LIT). Lastly, in section 5 we give some
final remarks.

\section{Method: Relativistic Kinetic Theory}

The Boltzmann equation \cite{Lichnerowicz} for a mixture of two non-reacting
dilute species in thermal local equilibrium reads,\begin{equation}
v_{(i)}^{\alpha}f_{(i),\alpha}=\sum_{j}J_{(ij)}\label{eq:1}\end{equation}
where the collisional term is given by \cite{CERCIGNANI},\begin{equation}
\sum_{j}J_{(ij)}=\sum_{j}\int\left(f'_{(i)}f'_{(j)}-f_{(i)}f_{(j)}\right)F_{(ij)}\Sigma_{(ij)}d\Omega_{(ji)}dv_{(j)}^{*}.\label{eq:2}\end{equation}
Here, $F_{(ij)}$, $\Sigma_{(ij)}$ and $d\Omega_{(ji)}$ denote the
invariant flux, the invariant differential elastic cross-section and
the element of solid angle that characterizes a binary collision between
particles of the same species as well as those between different species.
The differential $dv_{(i)}^{*}$ stands for $\frac{d^{3}v_{(i)}}{v_{(i)}^{4}}$
, which is also an invariant. The cross-section $\Sigma_{(ij)}$ has
special symmetries \cite{Marrot,Taub,Teo H Weert-Leeuwen-Groot} that
guarantee the existence of inverse collisions such that the principle
of microscopic reversibility and therefore an H theorem are satisfied. 

It is important to notice that the molecular velocity $v_{(i)}^{\mu}$
is measured by an observer in an arbitrary frame, in which the four-velocity
of the fluid is represented by $U^{\mu}$. This frame is called the
laboratory frame. 

From Eq. (\ref{eq:1}) one can obtain the balance equations by multiplying
it by the collisional invariants namely, the rest mass $m_{(i)}$,
the four-momentum $m_{(i)}v_{(i)}^{\mu}$ and then integrating over
the velocities $dv_{(i)}^{*}$. The complete set of equations can
be found in Ref. \cite{Val-1}. In this work we only need them at
lowest order in the gradients. This is accomplished through the use
of the Chapman-Enskog \cite{Ana 2008,Chapman-Cowling} method of solution
to Eq. (\ref{eq:1}). The particle number density conservation for
each species reads as,\begin{equation}
n_{(i)}U_{;\alpha}^{\alpha}+U^{\alpha}n_{(i),\alpha}=0,\label{eq:3}\end{equation}
the momentum balance as\begin{equation}
\tilde{\rho}U^{\mu}U_{;\mu}^{\beta}+h^{\beta\nu}p_{,\nu}=0\label{eq:4}\end{equation}
and finally, the energy conservation as\begin{equation}
nU^{\nu}e_{,\nu}=-pU_{;\mu}^{\mu}.\label{eq:5}\end{equation}
Here,\begin{equation}
n=n_{(i)}+n_{(j)}\end{equation}
is the total particle density, and\begin{equation}
\tilde{\rho}=\sum_{i}m_{(i)}n_{(i)}G\left(z_{(i)}\right)=\tilde{\rho}_{(1)}+\tilde{\rho}_{(2)},\label{eq:6}\end{equation}
with \begin{equation}
G\left(z_{(i)}\right)=\frac{\mathcal{K}_{3}\left(\frac{1}{z_{(i)}}\right)}{\mathcal{K}_{2}\left(\frac{1}{z_{(i)}}\right)},\label{eq:7}\end{equation}
and $h^{\beta\alpha}=g^{\beta\alpha}+c^{-2}U^{\beta}U^{\alpha}$ is
a projector in the direction orthogonal to $U^{\alpha}$. Here $\mathcal{K}_{n}\left(\frac{1}{z_{(i)}}\right)$
is the modified Bessel function of the second kind for the integer
$n$. 

We now expand Eq. (\ref{eq:1}) using the well-known Chapman-Enskog
series \cite{Ana 2008,Chapman-Cowling} up to first order in the gradients
namely,\begin{equation}
f_{(i)}=f_{(i)}^{(0)}\left(1+\phi_{(i)}\right),\label{eq:8}\end{equation}
where $f_{(i)}^{(0)}$ is the local equilibrium distribution namely,
Jüttner's distribution \cite{JUTTNER,Chacon Dagdug Morales} which
is given by, \begin{equation}
f_{(i)}^{(0)}=\frac{n_{(i)}}{4\pi c^{3}z_{(i)}\mathcal{K}_{2}\left(\frac{1}{z_{(i)}}\right)}\text{ exp}\left(\frac{U^{\beta}v_{(i)\beta}}{z_{(i)}c^{2}}\right),\label{eq:9}\end{equation}
with $z_{(i)}=\frac{k_{B}T}{m_{(i)}c^{2}}$ . Moreover, we must now
introduce the functional hypothesis namely $f_{(i)}\left(x^{\alpha},v_{(i)}^{\alpha}|n_{(i)},U^{\alpha},T\right)$,
implying that the representation chosen is defied by the locally conserved
variables $n_{(i)}$, $U^{\beta}$ and $T$. We remind the reader
that this assumption constitutes one possibility of extracting from
the manifold of the possible solutions of the Boltzmann equations
those which are consistent with the hydrodynamics of the fluid Ref.
\cite{Uhlenbeck} (also known in the literature as Hilbert's paradox).

From now on, we will develop all the calculations in the local co-moving
frame, where the spatial components of the hydrodynamical four-velocity
vanish, i.e. $U^{m}=0$. This frame has the advantage that allows
us to isolate the purely kinetic effects of the motion of the particles
from the convective effects \cite{Maxwell,Brush,Ana}. Then we can
transform all the quantities measured in such frame to an arbitrary
one with four-velocity $U^{\beta}$ through a Lorentz transformation.
Indeed, denoting by $\mathcal{L}_{\nu}^{\mu}$ the Lorentz transformation,
the molecular four-velocity in a moving frame reads as,\begin{equation}
v_{(i)}^{\mu}=\mathcal{L}_{\nu}^{\mu}K_{(i)}^{\nu}.\label{eq:10}\end{equation}
Here $K_{(i)}^{\nu}$ is the four-velocity in the local co-moving
frame. In the classical framework it is precisely the definition of
the well known peculiar or thermal velocity \cite{Chapman-Cowling}. 

The definition of the dissipative mass flux, heat flux and viscous
tensor are established when we obtain the complete set of transport
equations (see Ref. \cite{Val-1}) in the local co-moving frame. They
are given by, \begin{equation}
J_{(i)}^{m}=m_{(i)}\int K_{(i)}^{m}f_{(i)}dK_{(i)}^{*},\label{eq:11}\end{equation}
 \begin{equation}
q_{(i)}^{m}=m_{(i)}c^{2}\int\gamma_{k_{(i)}}K_{(i)}^{m}f_{(i)}dK_{(i)}^{*}\label{eq:12}\end{equation}
and\begin{equation}
\pi_{(i)}^{mn}=m_{(i)}\int K_{(i)}^{m}K_{(i)}^{n}f_{(i)}dK_{(i)}^{*}\label{eq:13}\end{equation}
respectively. Here $\gamma_{k_{(i)}}=\left(1-\frac{k_{(i)}^{2}}{c^{2}}\right)^{-1/2}$.

After some algebra, the expansion of the Boltzmann equation (\ref{eq:1})
with the help of Eqs. (\ref{eq:8}), (\ref{eq:3}), (\ref{eq:4}),
(\ref{eq:5}) leads to the following equation,\begin{equation}
\begin{array}{c}
K_{(i)}^{m}\left\{ -\gamma_{k_{(i)}}\frac{1}{z_{(i)}c^{2}\tilde{\rho}}p_{,m}+\left(\ln n_{(i)}\right)_{,m}+\left[1+\frac{1}{z_{(i)}}\left(\gamma_{k_{(i)}}-G\left(z_{(i)}\right)\right)\right]\left(\ln T\right)_{,m}\right\} \\
+\frac{1}{z_{(i)}c^{2}}\left(\overset{\text{\textdegree}}{K_{(i)}^{m}K_{(i)n}}\right)U_{;m}^{n}+\tau_{(i)}U_{;m}^{m}\\
=\left[C\left(\phi_{(i)}\right)+C\left(\phi_{(i)}+\phi_{(j)}\right)\right],\end{array}\label{eq:14}\end{equation}
where, \begin{equation}
K_{(i)}^{m}K_{(i)n}=\left(\overset{\text{\textdegree}}{K_{(i)}^{m}K_{(i)n}}\right)+\tau_{(i)}\delta_{n}^{m}.\label{eq:14.1}\end{equation}
Here $\tau_{(i)}$ corresponds to the trace of the tensor $K_{(i)}^{m}K_{(i)n}$
and it is related to the dynamic pressure, while $\left(\overset{\text{\textdegree}}{K_{(i)}^{m}K_{(i)n}}\right)$
denotes the symmetric traceless part. The collisional linearized kernels
are\begin{equation}
\left[C\left(\phi_{(i)}\right)+C\left(\phi_{(i)}+\phi_{(j)}\right)\right]=\sum_{i}\int\cdots\int f_{(i)}^{(0)}f_{(j)}^{(0)}\left(\phi_{(i)}\text{\textasciiacute}+\phi_{(j)}\text{\textasciiacute}-\phi_{(i)}-\phi_{(j)}\right)F_{(ij)}\Sigma_{(ij)}d\Omega_{(ji)}dK_{(j)}^{*}.\label{eq:14.2}\end{equation}

The form of Eq. (\ref{eq:14}) is crucial in order to identify the
thermodynamical forces. As we have said, the subtle decision of how
it can be rearranged has been studied in a previous work \cite{Val-3}
by addressing the necessity of complying with the ORR. This is accomplished
through the introduction of a pseudo-flux which arises strictly from
relativistic considerations and is directly related to the term containing
the pressure gradient ($\sim p_{,m}$, recall that $p_{(i)}=n_{(i)}k_{B}T$).
Such a flux, we insist appears only in the relativistic kinetic theory
with $p_{,m}$ acting as its direct force and can be explained by
taking the averages of microscopic Lorentz deformations of spatial
cells and thus we may refer to it as a {}``volume flux''. This idea
has never been dealt within the literature before \cite{kremer,Groot-Leewen-Weert}
and has the advantage of allowing a clear definition of both the Soret
and Dufour effects in a relativistic framework.

\section{Linear Theory}

By following the arguments that we have discussed in the previous
section it is possible to rearrange the linearized Boltzmann equation
(\ref{eq:14}) as follows\cite{Val-3},

\begin{equation}
\begin{array}{c}
K_{(i)}^{m}\left\{ d_{m}+\frac{1}{z_{(i)}}\left(\gamma_{k_{(i)}}-G\left(z_{(i)}\right)\right)\frac{T_{,m}}{T}-\left(\gamma_{k_{(i)}}-G\left(z_{(i)}\right)\right)V_{(i)m}\right\} \\
+\frac{1}{z_{(i)}c^{2}}\left(\overset{\text{\textdegree}}{K_{(i)}^{m}K_{(i)n}}\right)U_{;m}^{n}+\tau_{(i)}U_{;m}^{m}\\
=\left[C\left(\phi_{(i)}\right)+C\left(\phi_{(i)}+\phi_{(j)}\right)\right],\end{array}\label{eq:15}\end{equation}
for species $i$. Recall that there is a similar equation for species
$j$. Here\begin{equation}
V_{m}\equiv V_{(i)m}=\frac{m_{(i)}}{m_{(j)}}V_{(j)m}=\frac{n_{(i)}m_{(i)}}{\tilde{\rho}}\frac{p_{,m}}{p_{(i)}}\label{eq:16}\end{equation}
represents a new relativistic thermodynamic pseudo-force $V_{(i)m}$
related with Lorentz contractions of the mean free path of the particles.
In Ref. \cite{Val-3} it has been clearly shown that the corresponding
transport coefficients satisfy the symmetries required by the Onsager
reciprocity relations. Notice also that here we have a relativistic
generalization of the diffusive force\begin{equation}
d_{m}=d_{m(i)}=-d_{m(j)}=n_{(j)}\left(\frac{m_{(j)}G\left(z_{(j)}\right)-m_{(i)}G\left(z_{(i)}\right)}{\tilde{\rho}}\right)\frac{p_{,m}}{p}+\frac{n}{n_{(i)}}\left(n_{i0}\right)_{,m},\end{equation}
with $n_{i0}=\frac{n_{(i)}}{n}$. In the non-relativistic limit one
recovers the usual expression \cite{Groot Mazur,On the validity of the Onsager relations...},\begin{equation}
d_{m}\rightarrow\frac{n_{(j)}}{\rho p}\left(m_{(j)}-m_{(i)}\right)\nabla p+\frac{n}{n_{(i)}}\nabla n_{i0}.\end{equation}

Now we proceed to the solution of Eq. (\ref{eq:15}) \cite{Ana 2008,Curtis},
\begin{equation}
\phi_{(i)}=-K_{(i)}^{m}A_{(i)}\frac{T_{,m}}{T}-\sum_{j}K_{(j)}^{m}B_{(j)}^{(i)}V_{m}-\sum_{j}K_{(j)}^{m}D_{(j)}^{(i)}d_{m}-L_{(i)n}^{m}U_{;m}^{n}.\label{eq:17}\end{equation}

Notice that when Eq. (\ref{eq:17}) is substituted into Eqs. (\ref{eq:11})
and (\ref{eq:12}) we obtain the heat and mass fluxes in which the
Soret and Dufour effects are clearly identified. Indeed, de Soret
coefficient is given by,

\begin{equation}
\sum_{(i)}\int f_{(i)}^{(0)}K_{(i)}^{n}K_{n(i)}A_{(i)}dK_{(i)}^{*}\end{equation}
and the Dufour by,\begin{equation}
\sum_{(i),(j)}\int f_{(i)}^{(0)}\frac{1}{z_{(i)}}\left(\gamma_{k_{(i)}}-G\left(z_{(i)}\right)\right)K_{(i)}^{n}K_{n(i)}D_{(j)}^{(i)}dK_{(i)}^{*}\end{equation}
which have been shown to be symmetric in Ref. \cite{Val-3}.

As it has been thoroughly discussed in Ref. \cite{Val-3}, the pseudo
force $V_{m}$ corresponds to a relative flux directly associated
with Lorentz deformations in the microscopic geometrical aspects of
the system. Such a flux which we have called a {}``volume flux''
is a novel quantity. Although in non-relativistic fluid dynamics the
concept of a volume flow dates back to Burnett \cite{Haro}, it has
been recently revived by Brenner \cite{Brenner,Brenner-1,Brenner-2},
but in our previous work \cite{Val-3} it arises strictly from relativistic
considerations. In this paper, such flux which we will denote as $J_{V}^{m}$
is precisely the conjugate of the pseudo force $V_{m}$. Also it is
important to underline the fact that both $J_{V}^{m}$ and $V_{m}$
vanish in the non-relativistic limit.

Now we will use Eq. (\ref{eq:17}) to find the entropy production
which, in analogy with the non-relativistic case is identified from
a balance equation of the form \begin{equation}
\frac{\partial}{\partial t}\left(ns\right)+\nabla\cdot\left(\vec{J}_{s}\right)=\sigma.\label{eq:17.1}\end{equation}

To accomplish this task, we consider the entropy four-flux in an arbitrary
frame\begin{equation}
S^{\mu}\equiv-k_{B}\sum_{(i)}\int v_{(i)}^{\mu}f_{(i)}\left(\ln f_{(i)}-1\right)dv_{(i)}^{*},\label{eq:17.2}\end{equation}
which we decompose as \cite{CERCIGNANI},\begin{equation}
S^{\mu}=aU^{\mu}+\phi^{\mu},\label{eq:17.3}\end{equation}
where $\phi^{\mu}$ is a four vector orthogonal to $U^{\mu}$. The
invariant $a$ can be thus expressed as\begin{equation}
a=-\frac{S^{\mu}U_{\mu}}{c^{2}}\label{eq:17.4}\end{equation}
or \begin{equation}
a=\frac{k_{B}}{c^{2}}\sum_{(i)}U_{\mu}\int v_{(i)}^{\mu}f_{(i)}\left(\ln f_{(i)}-1\right)dv_{(i)}^{*}.\label{eq:17.5}\end{equation}
Recalling the properties of invariants, we notice that, \begin{equation}
U^{\mu}v_{(i)\mu}=-\gamma_{k_{(i)}}c^{2}.\label{eq:17.6}\end{equation}
Moreover, using that $dv_{(i)}^{*}=d^{3}v_{(i)}/\gamma_{v_{(i)}}=d^{3}K_{(i)}/\gamma_{k_{(i)}}$,
we have that \begin{equation}
a=-k_{B}\sum_{(i)}\int f_{(i)}\left(\ln f_{(i)}-1\right)d^{3}K_{(i)},\label{eq:17.7}\end{equation}
which is readily identified with the local entropy density $s$ for
the mixture. In appendix A, it is shown that the entropy four-flux
defined in Eq. (\ref{eq:17.2}) satisfies a balance equation of the
form\begin{equation}
S_{;\mu}^{\mu}=\sigma\label{eq:17.8}\end{equation}
 where $\sigma$ is the entropy production given by\begin{equation}
\sigma=-k_{B}\sum_{(i),(j)}\int J\left(f_{(i)}f_{(j)}\right)\ln f_{(i)}dv_{(i)}^{*}.\label{eq:17.9}\end{equation}

In the next section we will provide thermodynamical content to Eq.
(\ref{eq:17.9}). Indeed, what we will show is that $\sigma$ may
be related to Eq. (\ref{eq:0}) only when $f_{(i)}$ is written in
terms of the state variables.

\section{Derivation of $\sigma$}

In order to carry out the program outlined in the previous section,
we start with the property that the Boltzmann equation is a relativistic
invariant. We next choose the co-moving frame for performing the calculations
which will be carried out only to first order in gradients. For this
purpose we analyze Eq. (\ref{eq:17.9}) expanded with the help of
Eq. (\ref{eq:8}). Thus,\begin{equation}
\sigma=-k_{B}\sum_{(i),(j)}\int\cdots\int f_{(i)}^{(0)}f_{(j)}^{(0)}\left(\phi_{(i)}\text{\textasciiacute}+\phi_{(j)}\text{\textasciiacute}-\phi_{(i)}-\phi_{(j)}\right)\ln\left[f_{(i)}^{(0)}\left(1+\phi_{(i)}\right)\right]F_{(ij)}\Sigma_{(ij)}d\Omega_{(ji)}dK_{(j)}^{*}dK_{(i)}^{*}.\end{equation}
Now we expand $\ln\left[f_{(i)}^{(0)}\left(1+\phi_{(i)}\right)\right]$
in a neighborhood of $\phi_{(i)}<<1$, \begin{equation}
\ln\left[f_{(i)}^{(0)}\left(1+\phi_{(i)}\right)\right]\simeq\ln f_{(i)}^{(0)}+\phi_{(i)}+O\left(\phi_{(i)}\right)^{2},\label{eq:19-1}\end{equation}
so we have,\begin{equation}
\sigma=-k_{B}\sum_{(i),(j)}\int\cdots\int f_{(i)}^{(0)}f_{(j)}^{(0)}\left(\phi_{(i)}\text{\textasciiacute}+\phi_{(j)}\text{\textasciiacute}-\phi_{(i)}-\phi_{(j)}\right)\left(\ln f_{(i)}^{(0)}+\phi_{(i)}\right)F_{(ij)}\Sigma_{(ij)}d\Omega_{(ji)}dK_{(j)}^{*}dK_{(i)}^{*}.\end{equation}
Noticing that $\ln f_{(i)}^{(0)}$ is a combination of all collisional
invariants, all integrals associated to it vanish. Whence\begin{equation}
\sigma=-k_{B}\sum_{(i),(j)}\int\cdots\int f_{(i)}^{(0)}f_{(j)}^{(0)}\left(\phi_{(i)}\text{\textasciiacute}+\phi_{(j)}\text{\textasciiacute}-\phi_{(i)}-\phi_{(j)}\right)\phi_{(i)}F_{(ij)}\Sigma_{(ij)}d\Omega_{(ji)}dK_{(j)}^{*}dK_{(i)}^{*},\end{equation}
which, with the help of Eq. (\ref{eq:14.2}), is easily written as
\begin{equation}
\sigma=-k_{B}\sum_{(i),(j)}\int\left[C\left(\phi_{(i)}\right)+C\left(\phi_{(i)}+\phi_{(j)}\right)\right]\phi_{(i)}dK_{(i)}^{*}.\label{eq:20-1}\end{equation}
Finally, by using Eq. (\ref{eq:14}) we get,\begin{eqnarray}
\sigma & = & -k_{B}\sum_{(i)}\int f_{(i)}^{(0)}\left[K_{(i)}^{m}\left\{ d_{m}+\frac{1}{z_{(i)}}\left(\gamma_{k_{(i)}}-G\left(z_{(i)}\right)\right)\frac{T_{,m}}{T}-\left(\gamma_{k_{(i)}}-G\left(z_{(i)}\right)\right)V_{(i)m}\right\} \right.\nonumber \\
 &  & \left.+\frac{1}{z_{(i)}c^{2}}\left(\overset{\text{\textdegree}}{K_{(i)}^{m}K_{(i)n}}\right)U_{;m}^{n}+\tau_{(i)}U_{;m}^{m}\right]\phi_{(i)}dK_{(i)}^{*}.\label{eq:22-1}\end{eqnarray}
Where $\phi_{(i)}$ is given by Eq. (\ref{eq:17}). 

Carrying out the ensuing algebra (see appendix B) one arrives at,\begin{equation}
\frac{\sigma}{k_{B}}=-J^{m*}\left[d_{m}\right]-q^{m*}\left[\frac{T_{,m}}{T}\right]-J_{V}^{m}V_{m}-\frac{1}{k_{B}T}\pi_{n}^{m}U_{;m}^{n}-\tau U_{;m}^{m}.\label{eq:23}\end{equation}
In Eq. (\ref{eq:23}), we identify the vector fluxes namely, the diffusive
mass flux as, \begin{equation}
J^{m*}=\sum_{(i)}\frac{J_{(i)}^{m}}{m_{(i)}}=\sum_{(i)}\int K_{(i)}^{m}f_{(i)}^{(0)}\phi_{(i)}dK_{(i)}^{*},\label{eq:24.1}\end{equation}
the energy transport,\begin{equation}
q^{m*}=\frac{1}{k_{B}T}\sum_{(i)}\left(q_{(i)}^{m}-h_{(i)}\frac{J_{(i)}^{m}}{m_{(i)}}\right)\label{eq:24.2}\end{equation}
where \begin{equation}
h_{(i)}=\frac{k_{B}T}{z_{(i)}}G\left(z_{(i)}\right)\end{equation}
is the specific enthalpy, and the new ingredient, the {}``volume
flux'',\begin{equation}
J_{V}^{m}=\sum_{(i)}\left(J_{v(i)}^{m}-\frac{h_{E(i)}}{m_{(i)}c^{2}}\frac{J_{(i)}^{m}}{m_{(i)}}\right),\label{eq:24.3}\end{equation}
where $h_{E(i)}=\frac{\tilde{\rho}_{(i)}c^{2}}{n_{(i)}}$ depends
of the enthalpy through Eq. (\ref{eq:6}). The volume flux $J_{v(i)}^{m}$
is defined trough a corresponding balance equation \cite{Val-3}. 

On the other hand we have the tensor fluxes, \begin{equation}
\pi_{n}^{m}=\sum_{(i)}\pi_{(i)n}^{m}=\sum_{(i)}\int\overset{\text{\textdegree}}{\left(K_{(i)}^{m}K_{(i)n}\right)}f_{(i)}^{(0)}\phi_{(i)}dK_{(i)}^{*}\label{eq:25.1}\end{equation}
\begin{equation}
\tau=\sum_{(i)}\tau_{(i)}=\sum_{(i)}\int K_{(i)}^{n}K_{(i)n}f_{(i)}^{(0)}\phi_{(i)}dK_{(i)}^{*}.\label{eq:25.2}\end{equation}
It is important to notice that the structure of these fluxes is the
necessary required to obtain the canonical form for the entropy production
namely,\begin{equation}
\sigma=\sum_{i}J_{i}\odot X_{i},\label{eq:26}\end{equation}
where $\odot$ is the contraction to a scalar of fluxes with their
corresponding forces in accordance with Curie's theorem. Equation
(\ref{eq:26}) is in complete accordance with the Linear Irreversible
Thermodynamics. Notice that a completely equivalent argument can be
given for the other representation but since it has been carried out
in the literature \cite{CERCIGNANI} we omit it.

\section{Final Remarks}

In this work we have obtained the entropy production for a relativistic
binary mixture to first order in the gradients using the completely
new idea of {}``volume flux'' Ref. \cite{Val-3}. In such paper,
the {}``volume flux'' is produced with the next simple idea: Imagine
the motion of a single particle, then construct an imaginary volume
around it with radius $a$, where $a$ is the mean free path. Such
a volume would remain spherical in the non-relativistic scheme, but
because of the Lorentz contraction it would be deformed in the relativistic
framework. When we average these microscopic deformations per particle
we obtain the {}``volume flux''. Indeed, \textcolor{black}{when
we multiply Boltzmann's equation by the microscopic change in the
volume $a\gamma_{k_{(i)}}$, and then integrate over the velocities
$dK_{(i)}^{*}$ we find,}\begin{eqnarray}
\left(\int\gamma_{k(i)}K_{(i)}^{\alpha}f_{(i)}dK_{(i)}^{*}\right)_{,\alpha} & = & \int\gamma_{k_{(i)}}\left(J_{(ii)}+J_{(ij)}\right)dK_{(i)}^{*}\label{eq: vol flow}\\
 & = & \pi_{vol},\nonumber \end{eqnarray}
which is a balance equation for the change in the volume in the gas,
and defines, \begin{equation}
J_{v(i)}^{\alpha}=\int\gamma_{k(i)}K_{(i)}^{\alpha}f_{(i)}dK_{(i)}^{*}.\label{eq:jv}\end{equation}
Notice that in the non-relativistic limit, the right hand side vanishes,
implying that there is no such change in volume. On the other hand,
in the one-component limit $J_{v}^{\alpha}$ turns out to be a multiple
of the heat flux $q^{\alpha}$, indeed $\frac{q^{\alpha}}{k_{B}T}=\frac{1}{z}J_{v}^{\alpha}$. 

The corresponding expression for the {}``volume flux'' is valid
only in the co-moving frame. This however has no restriction since
to obtain the same quantity in an arbitrary frame one may simply resort
to the well-known Lorenz transformations.

The form of the fluxes here obtained in Eqs. (\ref{eq:24.1}) (\ref{eq:24.2})
(\ref{eq:24.3}) (\ref{eq:25.1}) and (\ref{eq:25.2}) is in accordance
with LIT and supports the definitions obtained in a previous work
\cite{Val-3}. Notice that, in the non-relativistic limit, the term
corresponding to the volume flux $J_{V}^{m}$ in Eq. (\ref{eq:23})
or (\ref{eq:vol flux}) vanishes because \begin{equation}
\gamma_{k_{(i)}}-G\left(z_{(i)}\right)\rightarrow0,\end{equation}
so we recover the classical expression for the entropy production
\cite{Chapman-Cowling}.

The generalization of the Soret and Dufour effects found in a previous
work Ref. \cite{Val-3} are now formally sustained by Eq. (\ref{eq:23}).
Their corresponding coefficients are easily found when Eq. (\ref{eq:24.1})
and (\ref{eq:24.2}) are expanded using the Chapman-Enskog method.
Additionally we have two more new coefficients related with $V^{m}$
whose physical meaning remains to be studied.

We would like to emphasize at this stage that the entropy production
as defined here and in general, in LIT is unfortunate \cite{entropy}.
Logically speaking it has no meaning since entropy as any other state
variable such as energy, pressure, volume, etc, can not be {}``produced''.
It is a pity that the original concept of uncompensated heat defined
as $T\sigma$ as originally introduced by Clausius Ref. \cite{Clausius}
has not been kept. Uncompensated heat is the energy that arises in
any thermodynamic process due to dissipative effects, and further
what one can measure in the laboratory is heat, not entropy. 
\begin{acknowledgments}
The authors wish to thank Alfredo Sandoval-Villalbazo for his helpful
comments and Universidad Iberoamericana Ciudad de Mexico for hosting
part of this work. One of us, V. M. acknowledges CONACyT for financial
support under scholarship number 203111.
\end{acknowledgments}

\section*{Appendix A}

Let \begin{equation}
S^{\mu}\equiv-k_{B}\sum_{(i)}\int v_{(i)}^{\mu}f_{(i)}\left(\ln f_{(i)}-1\right)dv_{(i)}^{*}\label{eq:A1}\end{equation}
be the entropy four-flux. Then,\begin{equation}
S_{;\mu}^{\mu}=\sigma\label{eq:A2}\end{equation}
 where\begin{equation}
\sigma=-k_{B}\sum_{(i),(j)}\int J\left(f_{(i)}f_{(j)}\right)\ln f_{(i)}dv_{(i)}^{*}\geq0\label{eq:A3}\end{equation}
and $J\left(f_{(i)}f_{(j)}\right)$ is defined in Eq. (\ref{eq:2}).

Let us start with the four-divergence of Eq. (\ref{eq:A1}),\begin{eqnarray}
S_{;\mu}^{\mu} & = & \left(-k_{B}\sum_{(i)}\int v_{(i)}^{\mu}f_{(i)}\left(\ln f_{(i)}-1\right)dv_{(i)}^{*}\right)_{;\mu}\label{eq:A4}\\
 & = & -k_{B}\sum_{(i)}\int\left[v_{(i)}^{\mu}f_{(i)}\left(\ln f_{(i)}-1\right)\right]_{;\mu}dv_{(i)}^{*}\nonumber \\
 & = & -k_{B}\sum_{(i)}\int\left[\left(v_{(i)}^{\mu}f_{(i)}\ln f_{(i)}\right)_{;\mu}-\left(v_{(i)}^{\mu}f_{(i)}\right)_{;\mu}\right]dv_{(i)}^{*}\nonumber \\
 & = & -k_{B}\sum_{(i)}\int\left[v_{(i);\mu}^{\mu}f_{(i)}\left(\ln f_{(i)}-1\right)+v_{(i)}^{\mu}f_{(i);\mu}\ln f_{(i)}\right]dv_{(i)}^{*}.\nonumber \end{eqnarray}
Now we substitute the Boltzmann Equation (\ref{eq:1}) in the second
term of the right hand side, which reads \begin{equation}
S_{;\mu}^{\mu}=-k_{B}\sum_{(i)}\int\left[v_{(i);\mu}^{\mu}f_{(i)}\left(\ln f_{(i)}-1\right)+\sum_{(j)}J\left(f_{(i)}f_{(j)}\right)\ln f_{(i)}\right]dv_{(i)}^{*},\label{eq:A5}\end{equation}
then\begin{equation}
-k_{B}\sum_{(i)}\int v_{(i);\mu}^{\mu}f_{(i)}\left(\ln f_{(i)}-1\right)dv_{(i)}^{*}=-k_{B}\sum_{(i),(j)}\int J\left(f_{(i)}f_{(j)}\right)\ln f_{(i)}dv_{(i)}^{*},\label{eq:A6}\end{equation}
so the entropy four-flux is given by\begin{equation}
S^{\mu}=-k_{B}\sum_{(i)}\int v_{(i)}^{\mu}f_{(i)}\left(\ln f_{(i)}-1\right)dv_{(i)}^{*}\label{eq:A7}\end{equation}
and the production term,\begin{equation}
\sigma=-k_{B}\sum_{(i),(j)}\int J\left(f_{(i)}f_{(j)}\right)\ln f_{(i)}dv_{(i)}^{*}.\label{eq:A8}\end{equation}

It is important to underline here that the identification of $S^{4}$
with the local entropy, $S^{m}$ with the entropy diffusive flux and
$\sigma$ with the non compensated heat is not complete until the
solution $f_{(i)}$ is determined. Indeed, while $f_{(i)}$ remains
unknown, none of these quantities can be taken as a function of thermodynamic
variables because they do not appear in $f_{(i)}$.

On the other hand, \begin{equation}
\sigma=-k_{B}\sum_{(i),(j)}\int J\left(f_{(i)}f_{(j)}\right)\ln f_{(i)}dv_{(i)}^{*}\label{eq:A9}\end{equation}
\begin{equation}
\sigma=-k_{B}\sum_{(i),(j)}\int\left(f'_{(i)}f'_{(j)}-f_{(i)}f_{(j)}\right)\ln f_{(i)}F_{(ij)}\sigma_{(ij)}d\Omega_{(ji)}dv_{(j)}^{*}dv_{(i)}^{*},\label{eq:A10}\end{equation}
which, by using the same transformations as those performed in the
proof the H theorem it can be written as\begin{equation}
\sigma=\frac{1}{4}k_{B}\sum_{(i),(j)}\int\left(f'_{(i)}f'_{(j)}-f_{(i)}f_{(j)}\right)\ln\frac{f'_{(i)}f'_{(j)}}{f_{(i)}f_{(j)}}F_{(ij)}\sigma_{(ij)}d\Omega_{(ji)}dv_{(j)}^{*}dv_{(i)}^{*}.\label{eq:A11}\end{equation}
Therefore, with the Klein's inequality,\begin{equation}
\left(f'_{(i)}f'_{(j)}-f_{(i)}f_{(j)}\right)\ln\frac{f'_{(i)}f'_{(j)}}{f_{(i)}f_{(j)}}\geq0,\label{eq:A12}\end{equation}
one obtains\begin{equation}
\sigma\geq0.\label{eq:A13}\end{equation}
We recall the reader that Eq. (\ref{eq:A13}) is valid for any exact
solution of the Boltzmann equation and has no inherent physical meaning
until one establishes the form of $f_{(i)}$.

\section*{Appendix B}

First recall Eq. (\ref{eq:22-1}), \begin{equation}
\sigma=-k_{B}\sum_{(i)}\int f_{(i)}^{(0)}\left[\begin{array}{c}
K_{(i)}^{m}\left\{ d_{m}+\frac{1}{z_{(i)}}\left(\gamma_{k_{(i)}}-G\left(z_{(i)}\right)\right)\frac{T_{,m}}{T}-\left(\gamma_{k_{(i)}}-G\left(z_{(i)}\right)\right)V_{(i)m}\right\} \\
+\frac{1}{z_{(i)}c^{2}}\left(\overset{\text{\textdegree}}{K_{(i)}^{m}K_{(i)n}}\right)U_{;m}^{n}+\tau_{(i)}U_{;m}^{m}\end{array}\right]\phi_{(i)}dK_{(i)}^{*},\end{equation}
so,\begin{equation}
\begin{array}{c}
\frac{\sigma}{-k_{B}}=\sum_{(i)}\int K_{(i)}^{m}\phi_{(i)}d^{3}K_{(i)}^{*}d_{m}+\sum_{(i)}\int K_{(i)}^{m}\frac{1}{z_{(i)}}\left(\gamma_{k_{(i)}}-G\left(z_{(i)}\right)\right)\phi_{(i)}dK_{(i)}^{*}\frac{T_{,m}}{T}\\
-\sum_{(i)}\int K_{(i)}^{m}\left(\gamma_{k_{(i)}}-G\left(z_{(i)}\right)\right)\phi_{(i)}dK_{(i)}^{*}V_{(i)m}\\
+\sum_{(i)}\int\frac{1}{z_{(i)}c^{2}}\left(\overset{\text{\textdegree}}{K_{(i)}^{m}K_{(i)n}}\right)\phi_{(i)}dK_{(i)}^{*}U_{;m}^{n}+\sum_{(i)}\int\tau_{(i)}\phi_{(i)}dK_{(i)}^{*}U_{;m}^{m}.\end{array}\end{equation}
Here, we identify the bilinear form: fluxes times forces structure.
The coefficient of the diffusive force is\begin{equation}
\frac{J_{(i)}^{m}}{m_{(i)}}=\int K_{(i)}^{m}\phi_{(i)}dK_{(i)}^{*}.\end{equation}
For the temperature gradient, the coefficient is \begin{eqnarray}
\sum_{(i)}\int K_{(i)}^{m}\frac{1}{z_{(i)}}\left(\gamma_{k_{(i)}}-G\left(z_{(i)}\right)\right)\phi_{(i)}dK_{(i)}^{*} & =\sum_{(i)} & \left[\right.\int K_{(i)}^{m}\frac{\gamma_{k_{(i)}}}{z_{(i)}}\phi_{(i)}dK_{(i)}^{*}\nonumber \\
 &  & -\frac{G\left(z_{(i)}\right)}{z_{(i)}}\int K_{(i)}^{m}\phi_{(i)}dK_{(i)}^{*}\left.\right]\nonumber \\
 & = & \frac{1}{k_{B}T}\left(q^{m}-h_{(i)}\frac{J_{(i)}^{m}}{m_{(i)}}\right),\end{eqnarray}
where $h_{(i)}=\frac{k_{B}T}{z_{(i)}}G\left(z_{(i)}\right)$ is the
specific enthalpy per species.

The coefficient of the new force $V_{(i)m}$ is given by\begin{eqnarray}
\sum_{(i)}\int K_{(i)}^{m}\left(\gamma_{k_{(i)}}-G\left(z_{(i)}\right)\right)\phi_{(i)}dK_{(i)}^{*} & = & \sum_{(i)}\left[\right.\int K_{(i)}^{m}\gamma_{k_{(i)}}\phi_{(i)}dK_{(i)}^{*}\nonumber \\
 &  & -G\left(z_{(i)}\right)\int K_{(i)}^{m}\phi_{(i)}dK_{(i)}^{*}\left.\right]\nonumber \\
 & = & \sum_{(i)}\left(J_{v(i)}^{m}-\frac{h_{E(i)}}{m_{(i)}c^{2}}\frac{J_{(i)}^{m}}{m_{(i)}}\right)\label{eq:vol flux}\end{eqnarray}
where $h_{E(i)}=\frac{\tilde{\rho}_{(i)}c^{2}}{n_{(i)}}$, and $J_{v}$
is defined in Eq. (\ref{eq:jv}).

\end{document}